\documentclass{article}
\usepackage{spconf}
\usepackage{hyperref}
\hypersetup{
    colorlinks=true,
    citecolor=blue,
    linkcolor=blue,
    filecolor=magenta,      
    urlcolor=blue,
}

\usepackage{spconf,amsmath}
\usepackage{adjustbox}
\usepackage{multirow}
\usepackage{multicol}
\usepackage{caption}
\usepackage{makecell}
\usepackage{hyperref}  
\usepackage{url}       
\usepackage[table]{xcolor}

\begin{document}

\title{ICASSP 2024 Speech Signal Improvement Challenge}
\name{\parbox{0.8\linewidth}{\centering
Nicolae-C\u{a}t\u{a}lin Ristea, Ando Saabas, Ross Cutler, Babak Naderi, Sebastian Braun, Solomiya Branets}}
\address{Microsoft Corp.}
\maketitle

\begin{abstract}
The ICASSP 2024 Speech Signal Improvement Grand Challenge is intended to stimulate research in the area of improving the speech signal quality in communication systems. This marks our second challenge, building upon the success from the previous ICASSP 2023 Grand Challenge. We enhance the competition by introducing a dataset synthesizer, enabling all participating teams to start at a higher baseline, an objective metric for our extended P.804 tests, transcripts for the 2023 test set, and we add Word Accuracy (WAcc) as a metric. We evaluate a total of 13 systems in the real-time track and 11 systems in the non-real-time track using both subjective P.804 and objective Word Accuracy metrics.
\end{abstract}

\begin{keywords}
Speech enhancement, SIG Challenge
\end{keywords}

\section{Introduction}
Audio communication systems such as remote collaboration systems (Microsoft Teams, Skype, Zoom, etc.), smartphones, and telephones are used by nearly everyone on the planet and have become essential tools for both work and personal usage. Despite the considerable effort invested until now, achieving a system comparable to face-to-face communication, particularly with the utilization of mainstream devices, remains a substantial journey.

The Speech Signal Improvement challenges are intended to stimulate research in the area of improving the speech signal quality in mainstream communication systems. The speech signal quality can be measured with SIG (signal) according to ITU-T P.835 test method, and is still a top issue in audio communication and conferencing systems \cite{cutler2023icassp}.
The objective of the ICASSP 2024 Speech Signal Improvement Grand Challenge is to once again unite researchers in addressing the high-impact topic of speech quality improvement. We provide the means to have their methodologies compared using a shared dataset and uniform evaluation procedures.

\section{Changes from the 2023 SIG Challenge}

In this challenge, we build upon the foundation laid during the ICASSP 2023 SIG Challenge \cite{cutler2023icassp}, incorporating several enhancements from insights gained in the earlier challenge.

\noindent \textbf{Data synthesizer.} In the previous year's challenge, it was noted that the leading teams achieved significant advancements through the use of well-designed dataset generators \cite{cutler2023icassp}. In order to elevate the starting point for all participating teams, we released a dataset synthesizer capable of closely replicating real-world settings \footnote{\url{https://github.com/microsoft/SIG-Challenge}}.

\noindent \textbf{P.804 objective metric.} In addition to existing objective metrics that estimate the subjective mean opinion score (MOS), we provide a new objective metric named SIGMOS, closely correlated with our extended P.804 tests \cite{naderi2023multi}. Notably, the SIGMOS metric assesses full-band audio, previously unexplored by comparable neural metrics. This allowed participants to evaluate, tune, and train their models without relying on subjective testing.

\begin{table*}[!t]
\begin{center}
\centerline{\includegraphics[width=1.0\linewidth]{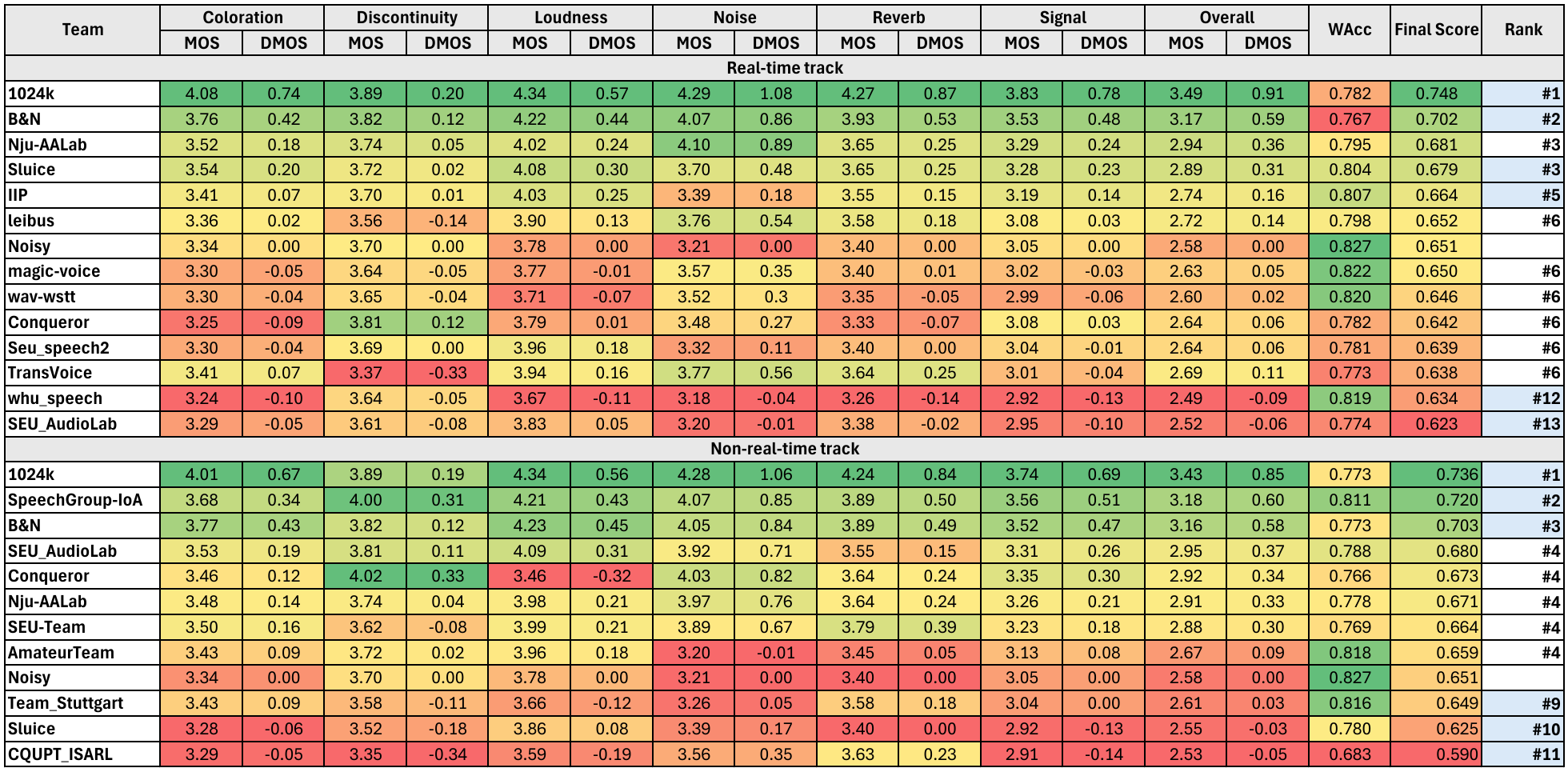}}
\caption{ICASSP 2024 Speech Signal Improvement Challenge results. We included MOS and differential MOS (DMOS) (for each score we subtract the corresponding Noisy score) for P.804 \cite{naderi2023multi} , WAcc and the Final Score with the corresponding rank. The statistically significant results at $p=0.05$ with respect to the next rank are highlighted in \textcolor{blue!60!white}{blue}.}
\label{fig_res}
\vspace{-1.0cm}
\end{center}
\end{table*}

\noindent \textbf{Improved evaluation procedure.}
In the 2023 challenge, our evaluation focused only on the SIG and OVRL (overall) subjective scores derived from P.804 tests \cite{naderi2023multi}. In the current challenge, we expand our evaluation to include not only the SIG and OVRL scores but also the Word Accuracy (WAcc) metric. This addition is particularly significant as WAcc exhibited surprisingly low correlation with P.804 results in \cite{cutler2023icassp}. By incorporating WAcc, we aim to provide a more comprehensive perspective on participants' results.

\noindent \textbf{Transcripts released.} 
To facilitate participants in independently assessing WAcc and evaluating their models, we released the transcripts for the public test set of the 2023 SIG challenge. This enables participants to directly compute and analyze WAcc, to evaluate, tune, and train their models.

\section{Challenge description}

\subsection{Blind dataset}
The blind set consists of 500 clips, each sourced from a distinct device, environment, and speaker. The clips were captured from both PCs and mobile devices, following the same methodology as described in the last challenge \cite{cutler2023icassp}. Additionally, a small percentage of clips were selected from the publicly available Common Voice data set.
To ensure diversity, the recordings were stratified to have an approximately uniform distribution for the impairment areas (e.g., noise, reverb, distortions), in accordance with labels given by expert listeners. The blind set language contains English, German, Dutch, French, and Spanish languages, with the majority of files (around 80\%) in English. The blind set was released near the end of the competition.

\subsection{Evaluation methodology}
Our evaluation is based on a subjective listening test, based on the P.804 standard \cite{naderi2023multi}, using the Amazon Mechanical Turk crowd-sourcing service. For quality control, we included both gold questions (clips where the expected answer for a scale is known ahead of time) and trapping questions (questions where rating clips are replaced with instructions to select a specific answer). Additionally, we added the WAcc rate using Azure Cognitive Services speech recognition to evaluate the intelligibility and speech preserving capacity. Overall, the challenge metric is:
\begin{equation*}
   \text{Final Score} = \frac{(\text{SIG}-1)/4 + (\text{OVRL}-1)/4 + \text{WAcc}}{3}
\end{equation*} 

We highlight that the blind set for both tracks is identical, the results being directly comparable.


\section{Results}

The challenge results are presented in Table~\ref{fig_res}. We conducted statistical testing (one-tailed related-sample t-test between systems adjacent to each other in the scoreboard, with no FWER correction) on the final scores to assess the significance of the observed differences at $p=0.05$. Notably, based on the final scores, the top 5 teams (1024k, B\&N, Nju-AALab, Sluice, IIP) emerge as winners in the real-time track. In the non-real-time track, the top 3 teams (1024k, SpeechGroup-IoA, B\&N) have statistically significant better results compared to their counterparts. For the subsequent 5 teams and noisy files, their results fall within the bounds of statistical significance when comparing performance across consecutive teams. We highlight that the proposed ranking could potentially mislead when teams share the same rank. While there may be no statistical difference between consecutive teams, the variance between non-consecutive teams could be statistically significant. For more details about the competition please refer to our challenge page\footnote{\url{https://aka.ms/sig-challenge}}.

We would like to thank all participants for their submissions and we hope that the challenge has served to move the state-of-the-art in signal enhancement forward.


\bibliographystyle{IEEEbib}
\bibliography{main}

\end{document}